\documentstyle[12pt]{article}

\renewcommand{\thesection}{\arabic{section}}
\renewcommand{\theequation}{\thesection.\arabic{equation}}
\begin{document}

\vskip 2.00cm

\title{\bf $AdS$ interpretation of two-point correlation function of 
QED}
\maketitle
\author{{{Sh. Mamedov\footnote{This talk was made on the basis of the 
work together with S. Parvizi}}
\footnote{Email: sh$_-$mamedov@yahoo.com 
\hspace{2mm} \&
\hspace{2mm}  shahin@theory.ipm.ac.ir }}}\\ 
\\{\small {\em 1. Institute for Studies in Theoretical Physics and
Mathematics (IPM),}} \\
{\small {\em P.O.Box 19395-5531, Tehran, Iran}}\\
{\small {\em 2. High Energy Physics Lab., Baku State University, }}\\
{\small {\em Z. Khalilov str.23, Baku 370148, Azerbaijan}}
\begin{abstract}
\noindent We have considered the two-point correlation
function of QED in worldline formalism. In position space it has been 
written in terms of heat kernel. This leads to
introducing the $K_1(x_i)$ function, which is related with the
bulk-to-boundary propagator of massless scalar field
and to reveal bulk-to-boundary propagator in the expression of photon 
polarization operator.
\end{abstract}
\hspace{15mm} \newline {\Large {\bf Introduction}} \vspace{4mm} 
\newline
\noindent

Study of correlation functions is the one of interesting topics of
$AdS$/CFT correspondence. In this connection it has meaning to
study the correlation
functions of realistic models, such as scalar and spinor QED in the 
$AdS$%
/CFT correspondence framework. Two-point correlation function of
electromagnetic field in QED is the photon polarization operator.
Worldline formalism on Schwinger parameter moduli space [1,2]
turned out useful for rewriting of two and three- point
correlation functions of free scalar field theory in terms of heat
kernel [3]. Using this approach $N$-photon amplitudes written in
terms of worldline formalism can be rewritten in terms of heat
kernel and can be interpreted in the $AdS$/CFT correspondence
language as well. So, we shall use this approach in order to
convert the photon polarization operator into the expressions
written in terms of bulk to boundary propagators in $AdS$
space-time.

\section{Photon polarization operator}

We shall use the expression of photon polarization operator in
scalar QED in Schwinger parametrization, which has coinciding
expression received both by direct calculation and in worldline
formalism [13, 14 ]:
\begin{equation}
\label{1}
\begin{array}{c}
\Pi \left( k_1,k_2\right) =-
\frac{\left( g\mu ^{\frac \epsilon 2}\right) ^2}{\left( 4\pi \right)
^{2-\frac \epsilon 2}}\int\limits_0^\infty \frac{d\tau }{\tau ^{\frac 
d2-1}}%
\int\limits_0^1d\alpha \ e^{\tau k_1k_2\alpha \left( 1-\alpha \right)
}\left[ \frac 2\tau \left( \delta \left( \alpha \right) -1\right) 
\epsilon
_1\cdot \epsilon _2+\left( 1-2\alpha \right) ^2\left( \epsilon _1\cdot
k_2\right) \left( \epsilon _2\cdot k_1\right) \right]  \\ \times
\int\limits_{-\infty }^\infty \frac{d^dz}{\left( 2\pi \right) 
^d}e^{i\left(
k_1+k_2\right) z}.
\end{array}
\end{equation}
Here $d$ is the dimension of space-time, $\epsilon _1,$ $\epsilon _2$ 
and $%
k_1,$ $k_2$ are polarization vectors and momenta of incoming and 
outgoing
photons, $\tau $ is the Schwinger parameter. Remark, the expression (1)
differs from two-point correlation function of free scalar field theory 
only
by additional square bracket [3]. In position space taking Gaussian
integrals over the momenta and derivatives, we obtain the following
expression of polarization operator%
\footnote{We have included constants into new one}:
\begin{equation}
\label{2}
\begin{array}{c}
\Pi \left( x_1,x_2\right) =-C\int\limits_0^\infty
\frac{d\tau }{\tau ^{\frac d2-1}}\int\limits_0^1d\alpha \
\int\limits_0^1d\beta \ \int\limits_{-\infty }^\infty d^dz\ \left( 
\frac \pi
{\tau \beta \alpha \left( 1-\alpha \right) }\right) ^{\frac 
d2}e^{-\frac{%
\left( z-x_1\right) ^2}{4\tau \beta \alpha \left( 1-\alpha \right) 
}}\left(
\frac \pi {\tau \left( 1-\beta \right) \alpha \left( 1-\alpha \right)
}\right) ^{\frac d2}e^{-\frac{\left( z-x_2\right) ^2}{4\tau \left( 
1-\beta
\right) \alpha \left( 1-\alpha \right) }} \\ \times \left[ \frac 2\tau
\left( \delta \left( \alpha \right) -1\right) \epsilon _1\cdot \epsilon
_2-\left( 1-2\alpha \right) ^2\frac 1{2\tau \left( 1-\beta \right) 
\alpha
\left( 1-\alpha \right) }\epsilon _1\cdot \left( z-x_2\right) \frac 
1{2\tau
\beta \alpha \left( 1-\alpha \right) }\epsilon _2\cdot \left( 
z-x_1\right)
\right] .
\end{array}
\end{equation}
We can write (2) in terms of the heat kernel $\left( \frac 1{4\pi 
t}\right)
^{\frac d2}\ e^{-\frac{\left( x-z\right) ^2}{4t}}=\left\langle x\left|
e^{t\Box }\right| z\right\rangle $ and make change of variables: $\rho
_1=\rho \left( 1-\beta \right) ,\ \rho _2=\beta \rho $ and $t=4\tau 
\rho
\beta \left( 1-\beta \right) \alpha \left( 1-\alpha \right) :$
\begin{equation}
\label{3}
\begin{array}{c}
\Pi \left( x_1,x_2\right) =-C^{\prime }\int\limits_0^\infty
\frac{dt}{t^{\frac d2-1}}\int\limits_0^1d\alpha \ \left[ 4\alpha \left(
1-\alpha \right) \right] ^{\frac d2-2}\int\limits_0^1d\beta \ \left( 
\rho
_1\rho _2\right) ^{\frac d2-2}\rho ^{-\left( \frac d2-2\right)
}\int\limits_{-\infty }^\infty d^dz\ \left\langle x_1\left| e^{\frac 
t{4\rho
_1}\Box }\right| z\right\rangle  \\ \times \left\langle x_2\left| 
e^{\frac
t{4\rho _2}\Box }\right| z\right\rangle \left[ \frac{8\rho _1\rho 
_2\alpha
\left( 1-\alpha \right) }{t\rho }\left( \delta \left( \alpha \right)
-1\right) \epsilon _1\cdot \epsilon _2-\left( 1-2\alpha \right) 
^2\frac{%
4\rho _1\rho _2}{t^2}\epsilon _1\cdot \left( z-x_2\right) \epsilon 
_2\cdot
\left( z-x_1\right) \right] .
\end{array}
\end{equation}
We can insert into (3) $\Gamma \left( s\right) $ function 
representations
for $1$:
\begin{equation}
\label{4}1=\frac 1{\Gamma \left( \frac d2\right) }\int\limits_0^\infty 
d\rho
\rho ^{\frac d2-1}e^{-\rho }.
\end{equation}

Of course, in terms, which contains different degrees of $\rho
_l$, we have to introduce $\Gamma \left( s\right) $ functions
having different value of argument. The change of integration
variables $\rho $ and $\beta $ into $\rho _1,\rho _2$ one's using
the equality $\int\limits_0^\infty \rho d\rho
\int\limits_0^1d\beta =\int\limits_0^\infty d\rho
_1\int\limits_0^\infty d\rho _2$ is turned out necessary for next
step. In this variables $\Pi \left( x_1,x_2\right) $ has got more
symmetric form:
\begin{equation}
\label{5}
\begin{array}{c}
\Pi \left( x_1,x_2\right) =\int\limits_0^\infty
\frac{dt}{t^{\frac d2+1}}\ \int\limits_{-\infty }^\infty
d^dz\int\limits_0^\infty d\rho _1\rho _1^{\frac d2-1}e^{-\rho
_1}\left\langle x_1\left| e^{\frac t{4\rho _1}\Box }\right| 
z\right\rangle
\int\limits_0^\infty d\rho _2\rho _2^{\frac d2-1}e^{-\rho 
_2}\left\langle
x_2\left| e^{\frac t{4\rho _2}\Box }\right| z\right\rangle \  \\ \times
\left[ -\frac{C_1}{\Gamma \left( \frac d2+1\right) }t\epsilon _1\cdot
\epsilon _2+\frac{C_2}{\Gamma \left( \frac d2\right) }\epsilon _1\cdot
\left( z-x_2\right) \epsilon _2\cdot \left( z-x_1\right) \right] .
\end{array}
\end{equation}
Here we have taken into account $\rho _1+\rho _2=\rho $ in the exponent 
and
have included integrals over the $\alpha $ into constants $C_{1,2}$%
\footnote{We suppose $d\geq 2$ for convergency of these integrals}:%
$$
\begin{array}{c}
C_1=4^{\frac d2-1}C^{\prime }\int\limits_0^1d\alpha \ \left[ \alpha 
\left(
1-\alpha \right) \right] ^{\frac d2-1}\left( \delta \left( \alpha 
\right)
-1\right) , \\
C_2=4^{\frac d2-2}C^{\prime }\int\limits_0^1d\alpha \ \left[ \alpha 
\left(
1-\alpha \right) \right] ^{\frac d2-1}\left( 1-2\alpha \right) ^2.
\end{array}
$$
Following [3] we have separated the integrals over the $\rho _i$
and denote them by $K_1\left( x_i,z,t\right) $ function:
\begin{equation}
K_1\left( x_i,z,t\right) =\int\limits_0^\infty d\rho _i\rho _i^{\frac
d2-1}e^{-\rho _i}\left\langle x_i\left| e^{\frac t{4\rho _i}\Box 
}\right|
z\right\rangle .
\end{equation}
As was shown in [3], identifying the variable $t$ with the radius
of $d$ -dimensional sphere $z_0$ $\left( t=z_0^2\right) $ the
function $K_1\left( x_i,z,t\right) $ obeys the $d+1$-dimensional
Klein-Gordon equation with zero mass:
\begin{equation}
\label{6}\left[ -z_0^2\partial _{z_0}^2+\left( d-1\right) z_0\partial
_{z_0}-z_0^2\Box \right] K_1\left( x,z,t\right) =0,
\end{equation}
where $\Box $ is the $d$-dimensional Laplacian in the direction $
\overrightarrow{z}$. That means that function $K_1\left(
x_i,z,t\right) $ is the bulk to boundary propagator of massless
scalar or vector field in the $d+1$-dimensional $AdS$ space-time.
Now we can write (5) in terms of this propagator in more suitable
form for $AdS$/CFT interpretation:
\begin{equation}
\label{7}
\begin{array}{c}
\Pi \left( x_1,x_2\right) =\int\limits_0^\infty
\frac{dt}{t^{\frac d2+1}}\int\limits_{-\infty }^\infty d^dz\ K_1\left(
x_1,z,t\right) K_1\left( x_2,z,t\right)  \\ \times \ \left[ 
-\frac{C_1}{%
\Gamma \left( \frac d2+1\right) }\epsilon _1\cdot \epsilon _2t+C_2\frac
1{\Gamma \left( \frac d2\right) }\epsilon _1\cdot \left( z-x_2\right)
\epsilon _2\cdot \left( z-x_1\right) \right] .
\end{array}
\end{equation}
Comparing this expression for $\Pi \left( x_1,x_2\right) $ with the
two-point correlation function $\Gamma \left( x_1,x_2\right) $ for free
scalar field theory, we find additional square bracket factor in our 
case,
which should be replaced by $t^3$ for last one. Thus, the photon
polarization operator is shown in terms of bulk to boundary propagator 
$%
K_1\left( x,z,t\right) $ of massless field. For spinor QED case, when 
we
have spinor particles in the loop, the photon polarization operator has 
form
minor changing in its Schwinger parameter expression (1) for scalar 
loop
[1,2]:
\begin{equation}
\label{8}
\begin{array}{c}
\Pi \left( k_1,k_2\right) =2
\frac{\left( g\mu ^{\frac \epsilon 2}\right) ^2}{\left( 4\pi \right)
^{2-\frac \epsilon 2}}\int\limits_0^\infty \frac{d\tau }{\tau ^{\frac 
d2-1}}%
\int\limits_0^1d\alpha \ e^{\tau k_1k_2\alpha \left( 1-\alpha \right)
}\left[ \frac 2\tau \left( \delta \left( \alpha \right) -1\right) 
\epsilon
_1\cdot \epsilon _2\right.  \\ \left. -\left[ \left( 1-2\alpha \right)
^2-1\right] \left( \epsilon _1\cdot k_2\right) \left( \epsilon _2\cdot
k_1\right) +\left( \epsilon _1\cdot \epsilon _2\right) \left( k_1\cdot
k_2\right) \right] \int\limits_{-\infty }^\infty \frac{d^dz}{\left( 
2\pi
\right) ^d}e^{i\left( k_1+k_2\right) z}.
\end{array}
\end{equation}
This allows us to remake the formula (7) for spinor loop case (8):
\begin{equation}
\label{9}
\begin{array}{c}
\Pi \left( x_1,x_2\right) =2\int\limits_0^\infty
\frac{dt}{t^{\frac d2+1}}\int\limits_{-\infty }^\infty d^dz\ K_1\left(
x_1,z,t\right) K_1\left( x_2,z,t\right) \left[ C_1\frac{\epsilon 
_1\cdot
\epsilon _2}{\Gamma \left( \frac d2+1\right) }t\right.  \\ \left. 
+\frac
1{\Gamma \left( \frac d2\right) }\ \left[ C_2\epsilon _1\cdot \left(
z-x_2\right) \epsilon _2\cdot \left( z-x_1\right) -C_3\epsilon _1\cdot
\epsilon _2\left( z-x_1\right) \cdot \left( z-x_2\right) \right] 
\right] .
\end{array}
\end{equation}
Thus, we see from (7) and (9) two-point correlation functions of
vector field for scalar and spinor QED are expressed by means of
massless bulk-to-boundary propagator.  Since here we have studied
two-point correlation function of massless vector field ( photon
field ), the obtained result have expressed in terms of bulk to
boundary propagator of this field. If we multiply (7) to 2 and add
to (9), then we find good agreement with the result obtained in
[4]. This tells us that supersymmetry plays important role for matching 
correlation
functions in field theory with the $AdS$
supergravity ones.

\end{document}